\documentclass[12pt]{article}
\usepackage{latexsym,epsfig,amssymb, amsmath, amsfonts, units, cite, mcite}
\usepackage{enumerate}

\def\bfone{\relax{\rm 1\kern-.35em 1}}

\newcommand{\be}{\begin{equation}}
\newcommand{\ee}{\end{equation}}
\newcommand{\ben}{\begin{displaymath}}
\newcommand{\een}{\end{displaymath}}
\newcommand{\bea}{\begin{eqnarray}}
\newcommand{\eea}{\end{eqnarray}}

\newcommand{\eins}{\mbox{$1 \hspace{-1.0mm} \text{l}$}}
\newcommand{\bean}{\begin{eqnarray*}}
\newcommand{\eean}{\end{eqnarray*}}

\newcommand{\bbP}{\mathbb{P}}

\newcommand{\cR}{\mathcal{R}}

\newcommand{\cD}{\mathcal{D}}
\newcommand{\cG}{\mathcal{G}}
\newcommand{\cK}{\mathcal{K}}
\newcommand{\cL}{\mathcal{L}}
\newcommand{\cW}{\mathcal{W}}

\newenvironment{matr}[1]
{\left[ \begin{array}{{#1}}}{\end{array} \right]}

\makeatletter \@addtoreset{equation}{section} \makeatother

\begin{document}

\begin{titlepage}

\begin{flushright}
\small ~ \\
\end{flushright}

\bigskip

\begin{center}

\vskip 2cm

{\LARGE \bf A geometric bound on F-term inflation} \\

\vskip 1.0cm

{\bf Andrea Borghese, Diederik Roest, Ivonne Zavala } \\

\vskip 0.5cm

{\em Centre for Theoretical Physics,\\
University of Groningen, \\
Nijenborgh 4, 9747 AG Groningen, The Netherlands\\
{\small {\tt \{ a.borghese, d.roest, e.i.zavala \} @rug.nl}}} \\

\end{center}

\vskip 2cm

\begin{center} {\bf ABSTRACT}\\[3ex]

\begin{minipage}{13cm}
\small

We discuss a general bound on the possibility to realise inflation in any minimal supergravity with F-terms. 
The derivation crucially depends on the sGoldstini, the scalar field directions that are singled out by spontaneous 
supersymmetry breaking. The resulting bound involves both slow-roll parameters and the geometry of the K\"ahler manifold 
of the chiral scalars. We analyse the inflationary implications of this bound, and in particular discuss to what extent 
the requirements of single field and slow-roll can both be met in F-term inflation.

\end{minipage}

\end{center}

\vspace{2cm}

\vfill

\end{titlepage}


\tableofcontents

\section{Introduction}

Three decades after its inception, inflation remains our best theoretical candidate to describe the very early Universe 
\cite{Guth:1980zm, *Linde:1981mu}. It naturally explains the high degree of homogeneity at large scales in the present Universe, 
thus solving classical problems associated with e.g.~the horizon of CMB beyond patches of 1 degree and the nearly flat spatial geometry. 
In addition to this homogeneity, it also provides a compelling explanation for the small inhomogeneities in both the CMB and the LSS. 
The interpretation that these have originated from quantum fluctuations during inflation has been experimentally confirmed by the power 
spectrum of the CMB \cite{WMAP7}. As a result, we know the inflationary fluctuations are to a large extent Gaussian and almost scale invariant. 
In terms of  inflationary models, all observations so far are perfectly consistent with the simplest class of slow-roll and single field.
 The first constraint implies that the two slow-roll parameters,
 \begin{align}
    \epsilon \equiv \frac{\cG^{IJ} \, \cD_I V \, \cD_J V}{2 \, \ell_{p}^2 \, V} \,, \qquad
    \eta \equiv {\rm min.~eigenvalue~} \big(\frac{ \cG^{IK} \, \cD_K \cD_J V}{\ell_{p}^2 \, V} \big) \,,
 \end{align}
are both much smaller than unity\footnote{In what follows we  use $\ell_{p}^{2} = 1/M_{p}^{2} = 8 \pi G_{N}$.}. Indeed, the observed value of 
  \begin{align}
   n_s = 1 - 6 \epsilon + 2 \eta = 0.968 \pm 0.012 \,, \label{spectral}
 \end{align}
is consistent with percent level values for both slow-roll parameters. Future experiments such as Planck \cite{Planck} will measure the temperature
 anisotropies in far greater detail, and hence could observe deviations from this simple model, e.g.~by measuring non-Gaussianities.

In view of the phenomenological success of the inflationary hypothesis, it is natural to look for an embedding of this theory into a more 
fundamental theory of quantum gravity, such as string theory. Indeed in recent years a large research effort has been devoted to finding 
realisations of inflation in string theory. Despite a number of interesting and influential examples, this search is somewhat hampered by 
our limited knowledge of the contours of the playground: it remains unclear to this day which string theory compactifications are admissible, 
and what their resulting features are. 

A fruitful approach has been to limit oneself to a subset of all possibilities that we do understand. 
One example is D-brane inflation (see e.g.~\cite{DbraneIn} for a  review with several references), which has seen a lot of progress in the 
last years in the context of type IIB flux compactifications, 
where moduli stabilisation is under good theoretical control. Much effort has been put in computing the scalar potential of the D3-brane position, 
which is responsible to drive inflation. 
Another approach within type IIB flux compactifications has been to study modular inflation in the Large Volume scenario \cite{LV}. 
Here  the scalar potential responsible to drive inflation can be explicitely computed  and inflation realised \cite{CQ}.   

Another example is provided by the analysis of the inflationary properties of flux compactifications 
of type IIA string theory. Restricting oneself to Calabi-Yau compactifications with only standard NS-NS 3-form flux, R-R fluxes, 
D6-branes and O6-planes at large volume and small string coupling, one can stabilise the moduli at the classical level \cite{DeWolfe}. 
However, such constructions always satisfy a very simple and nevertheless strong lower bound on the first slow-roll parameter \cite{Hertzberg}: 
 \begin{align}
   \epsilon \geq {27} / {13} \,,
 \end{align}
violating the slow-roll assumption. Surprisingly, in order to derive this lower bound, only two of the total set of moduli fields had to be 
taken into account: one finds violation of the slow-roll condition already in the projection onto the two-dimensional plane spanned by the 
dilaton and the volume modulus. Based solely on the dynamics of these two fields, it has been argued that cosmological observations have ruled 
out geometric IIA compactifications. A possible way to circumvent this no-go theorem would be to replace the six-torus by negatively curved internal 
manifolds. 

In this paper we want to extend the above analyses in a different direction, namely that of minimal supergravity with an F-term scalar potential 
in terms of an arbitrary holomorphic superpotential. This is a very general class of models in which it is natural to embed inflation. 
The superpotential induces a potential energy for the K\"{a}hler fields, hence allowing for the possibility to realise inflation. Moreover, 
as the field content of minimal supergravity can contain an arbitrary number of chiral multiplets, it leaves room for all the subtleties of 
multi-field inflation, curvatons, isocurvature perturbations and non-Gaussianities, to name a few. Nevertheless, we will demonstrate that 
there is a very strong bound on such theories in order to satisfy \eqref{spectral}.

Instead of taking on this problem head-on, or statistically sample a large number of possibilities as in \cite{mcallister}, we 
derive an analytic bound by employing a simplification analogous to the two-dimensional projection of \cite{Hertzberg}. However, in our case 
the only directions out of all K\"{a}hler fields that are singled out are the so-called sGoldstini directions. These are the scalar partners 
of the would-be Goldstino that are eaten up by the gravitino in the process of supersymmetry breaking. 
Therefore, it is supersymmetry breaking that dynamically determines a number of preferred directions in moduli space. 

It has been shown in various supergravity contexts that the sGoldstini directions are very efficient in tracing possible scalar instabilities 
\cite{GR1, *GR2, *Louis}. For this reason, one can use the sGoldstini directions to derive an upper bound on the second slow-roll parameter $\eta$.  This slow-roll condition on F-term supergravity was discussed in \cite{Covi}. Note that it does not require the sGoldstini and inflaton directions to coincide. We  demonstrate that additionally imposing the condition of
effective single field inflation leads to a much stronger bound. We discuss the inflationary implications of this bound and see to what extent it 
allows for e.g.~single field and slow-roll inflation. Intriguingly, we also find the necessity to introduce a negatively curved manifold as in
 \cite{Hertzberg}, but now as the scalar manifold spanned by the K\"{a}hler fields instead of the internal compactification manifold. We will argue that this rules out the possibility of small field inflation.

The organisation of this paper is as follows. In section 2 we derive the general bound on the inflationary slow-roll parameters for any
 F-term supergravity. Subsequently we analyse the inflationary implications of this bound in section 3. Section 4 contains our concluding remarks.

{\bf Note added:} Upon completion of this manuscript we became aware of the preprint \cite{Ana} where related issues regarding the possibility 
to realise inflation with only the sGoldstino field are discussed. We briefly comment on the relation to our findings in the conclusions. It would be 
interesting to compare both papers in more detail.

\section{Minimal supergravity with F-terms}

\subsubsection*{The Lagrangian}

The field content of ${\cal N}=1$ supergravity is given by a graviton $e_{\mu}{}^{a}$ and a gravitino $\psi_{\mu}$ coupled to $n$ 
chiral supermultiplets. Each of these is composed by a chiral spin-$1/2$ field $\chi$ and a complex field $\phi$. 
It has been shown that the $\phi^{i}$, $i = 1, \ldots n$ fields organise themselves in a 
K\"{a}hler-Hodge manifold. This geometric structure is a fundamental ingredient in building the theory. 

The Lagrangian is given by (modulo four fermion terms)
\begin{align*}
e^{-1} \, \cL = \cL_{\textsc{kin}} + \cL_{\textsc{f-m}} - V \, ,
\end{align*}
where
\begin{align} \label{Kinetic Lagrangian}
\cL_{\textsc{kin}} & = +\frac{1}{2 \ell_{p}^{2}} \, R - \frac{1}{2} \, \bar{\psi}_{\mu} \, \gamma^{\mu \nu \rho} \, \cD_{\mu} \psi_{\rho} - \cG_{i \bar{\jmath}} \, \partial_{\mu} \phi^{i} \, \partial^{\mu} \bar{\phi}^{\bar{\jmath}} \, + \\ 
& \quad + \frac{\ell_{p}^{2}}{2} \, \cG_{i \bar{\jmath}} \, \left( \bar{\chi}^{i} \, \gamma^{\mu} \cD_{\mu} \bar{\chi}^{\bar{\jmath}} + \bar{\chi}^{\bar{\jmath}} \, \gamma^{\mu} \cD_{\mu} \chi^{i} \right) + \frac{\ell_{p}^{2}}{\sqrt{2}} \, \cG_{i \bar{\jmath}} \, \left( \bar{\psi}_{\mu} \gamma^{\nu} \partial_{\nu} \bar{\phi}^{\bar{\jmath}} \, \gamma^{\mu} \, \chi^{i} + \text{h.c.} \right) \, . \nonumber
\end{align}
The gravitino $\psi_{\mu}$ is a Majorana spinor while $\chi^{i}$ is a left-handed spinor:
 \begin{align*}
 \bbP_{L} \, \chi^{i} = \tfrac12 \, (\eins + \gamma_{5}) \, \chi^{i} = \chi^{i} \,.
\end{align*}

The fermionic mass terms are given by
\begin{align} \label{Fermionic mass terms}
\cL_{\textsc{f-m}} & = + \frac{\ell_{p}^{2}}{2} \, e^{\ell_{p}^{2}  \frac{\cK}{2}} \, \cW \, \bar{\psi}_{\mu} \bbP_{R} \, \gamma^{\mu \nu} \psi_{\nu} 
+ \frac{\ell_{p}^{2}}{\sqrt{2}} \, e^{\ell_{p}^{2}  \frac{\cK}{2}} \, \cD_{i} \cW \, \bar{\psi}_{\mu} \, \gamma^{\mu} \chi^{i} -
 \frac{\ell_{p}^{2}}{2} \, e^{\ell_{p}^{2}  \frac{\cK}{2}} \, \cD_{i} \, \cD_{j} \cW \, \bar{\chi}^{i} \chi^{j} + \text{h.c.} \, ,
\end{align}
and the scalar potential is given by
\begin{align} \label{Scalar potential}
V = - 3 \, \ell_{p}^{2} e^{ \ell_{p}^{2} \cK} \, \cW \, \overline{\cW} + e^{ \ell_{p}^{2} \cK} \, \cG^{i \bar{\jmath}} \, \cD_{i} \cW \, \cD_{\bar{\jmath}} \overline{\cW} \, .
\end{align}
Every derivative is covariantised w.r.t.~K\"{a}hler transformations besides local Lorentz transformations. Whenever a 
derivative acts on a quantity with K\"{a}hler indices $(i, \bar{\imath})$ it needs to be further covariantised w.r.t.~diffeomorphisms 
on the K\"{a}hler manifold\footnote{More on our conventions and a detailed derivation of
(\ref{Kinetic Lagrangian})-(\ref{Scalar potential}) can be found in \cite{SUGRABOOK}.}. 

The Lagrangian is therefore fully specified by the following two quantities:
 \begin{itemize}
\item
$\cK = \cK(\phi^{i} , \bar{\phi}^{\bar{\imath}})$ is the K\"{a}hler potential and, by definition, the metric on the K\"{a}hler manifold is given by $\partial_{i} \partial_{\bar{\jmath}} \, \cK \equiv \cG_{i \bar{\jmath}}$. It has mass dimension two while the scalar fields $\phi^{i}$ are normalised to the Planck mass.
\item
$\cW = \cW(\phi^{i})$ is the holomorphic superpotential, which has mass dimension three.
\end{itemize}
The scalar potential (\ref{Scalar potential}) of any F-term supergravity is made up of two opposing contributions. The negative definite term, related to the superpotential itself, sets the AdS scale. In contrast, the positive definite term is related to the first covariant derivatives of the superpotential $\cD_{i} \cW$. The latter quantities are referred to as F-terms and play an essential role as the order parameter for supersymmetry breaking.

\subsubsection*{Scalar mass matrix}

The standard way of carrying out the analysis of this class of supergravity theories is by considering the following real combination\footnote{As one can see from (\ref{G potential}), the potential $\cG$ is ill-defined whenever $\cW = 0$. Stricly speaking, our analysis therefore no longer applies in this case. Nevertheless, we have explicitly checked that also when $\cW = 0$ the same conclusions, and in particular the bound \eqref{General bound}, still hold. An interesting example of such a model is \cite{Kallosh1, *Kallosh2}. We thank Renata Kallosh for correspondence on this point.}
\begin{align} \label{G potential} 
\cG = \cK + \ell_{p}^{-2} \, \ln \left| \ell_{p}^{3} \cW \right|^{2} \, .
\end{align}
This function is by construction K\"{a}hler invariant and hence one does not have to worry about covariantising derivatives w.r.t.~this 
kind of transformations. In terms of this function the scalar potential reads
\begin{align} \label{Scalar potential G}
V = \ell_{p}^{-4} \, e^{\ell_{p}^{2} \cG} \big(\ell_{p}^{2} \, \cG^{i \bar{\jmath}} \, \cG_{i} \, \cG_{\bar{\jmath}} - 3 \big) \, ,
\end{align}
where $\cG_{i}$ denotes the simple partial derivative of $\cG$ w.r.t.~$\phi^{i}$.
The first derivative is given by
\begin{align} \label{Gradient}
\partial_{i} V = \ell_{p}^{2} \, \cG_{i} \, V + \ell_{p}^{-2} \, e^{\ell_{p}^{2} \cG} \, \big( \cG_{i} + \cG^{j} \, \cD_{i} \cG_{j} \big) \, . 
\end{align}
The second derivatives are thus
\begin{align} \label{Hessian}
\cD_{i} \cD_{j} V & = \ell_{p}^{2} \, \big( \cG_{i} \, \cD_{j} V + \cG_{j} \, \cD_{i} V \big) + \ell_{p}^{2} \,\big( \cD_{i} \cG_{j} + \ell_{p}^{2} \, \cG_{i} \cG_{j} \big) \, V + \nonumber \\
& \quad + \ell_{p}^{-2} e^{\ell_{p}^{2} \cG} \, \big( 2 \, \cD_{(i} \cG_{j)} + \cG^{k} \, \cD_{i} \cD_{j} \cG_{k} \big) \, , \nonumber \\
 \cD_{\bar{\imath}} \cD_{j} V & = \ell_{p}^{2} \, \big( \cG_{\bar{\imath}} \, \cD_{j} V + \cG_{j} \, \cD_{\bar{\imath}} V \big) + \ell_{p}^{2} \,  \big( \cG_{\bar{\imath} j} - \ell_{p}^{2} \, \cG_{\bar{\imath}} \cG_{j} \big) \, V + \nonumber \\
& \quad + \ell_{p}^{-2} e^{\ell_{p}^{2} \cG} \, \big[ \cG_{\bar{\imath} j} + \big( \cD_{\bar{\imath}} \cG^{k} \big) \, \big( \cD_{j} \cG_{k} \big) - \cR_{j \bar{\imath} k \bar{l}} \, \cG^{k} \cG^{\bar{l}} \big] \, .
\end{align}
Using these derivatives we are able to construct the squared mass matrix for the scalar fields at any point in field space. It is given by
\begin{align} \label{Mass matrix}
m^{2 \, I}{}_{J} = \begin{matr}{cc} m^{2 \, i}{}_{j} & m^{2 \, i}{}_{\bar{\jmath}} \\ m^{2 \, \bar{\imath}}{}_{j} & m^{2 \, \bar{\imath}}{}_{\bar{\jmath}} \end{matr} =  \begin{matr}{cc} \cG^{i \bar{k}} \, \cD_{\bar{k}} \cD_{j} V &  \cG^{i \bar{k}} \, \cD_{\bar{k}} \cD_{\bar{\jmath}} V \\ \cG^{\bar{\imath} k} \, \cD_{k} \cD_{j} V & \cG^{\bar{\imath} k} \, \cD_{k} \cD_{\bar{\jmath}} V \end{matr} \, . 
\end{align}
where we have used the collective index $I = (i,\bar{\imath})$.

\subsubsection*{sGoldstino directions}

Spontaneous supersymmetry breaking is induced by $\cD_{i} \cW$. We consider a configuration of the theory in which supersymmetry is 
broken: $\cD_{i} \cW \neq 0$. We see from (\ref{Fermionic mass terms}) that the mixing between the gravitino and the chiral spin-$1/2$ 
fields is sourced exactly by the order parameter of supersymmetry breaking and is encoded in the term
\begin{align*}
\frac{\ell_{p}^{2}}{\sqrt{2}} \, e^{\ell_{p}^{2}  \frac{\cK}{2}} \, \cD_{i} \cW \, \bar{\psi}_{\mu} \, \gamma^{\mu} \chi^{i} = - 
\frac{1}{\ell_{p}} \,  \bar{\psi}_{\mu} \, \gamma^{\mu} \, \big( \bbP_{L} \zeta \big) \, ,
\end{align*}
where we have defined a linear combination of spin-$1/2$ fields
\begin{align} \label{Goldstino field}
\bbP_{L} \zeta = - \frac{\ell_{p}^{3}}{\sqrt{2}} \, e^{\ell_{p}^{2}  \frac{\cK}{2}} \, \cD_{i} \cW \, \chi^{i} \, .
\end{align}
This field is usually called the Goldstino. Indeed, it is possible to show that the dynamics of the gravitino can be disentangled from that of the spin-$1/2$ fields, by performing a supersymmetry transformation in which the supersymmetry parameter $\varepsilon$ is proportional to 
the Goldstino. Going to the so-called unitary gauge it is possible to eliminate from the spectrum the Goldstino. This is the analogue of the 
Higgs mechanism for spontaneous gauge symmetry breaking, often called super-Higgs mechanism (see for instance \cite{Cremmer:1982wb}). 
The missing degrees of freedom are absorbed by the gravitino.

We now consider the supersymmetry variation of the Goldstino field. Apart from terms involving fermions, it is given by
\begin{align*}
\delta \big(\bbP_{L} \zeta \big) & = - \frac{\ell_{p}^{3}}{2} \, e^{\ell_{p}^{2}  \frac{\cK}{2}} \, \cD_{i} \cW \, \frac{1}{\ell_{p}} \gamma^{\mu} \, \partial_{\mu} \phi^{i} \big( \bbP_{R} \varepsilon \big) + \frac{\ell_{p}^{2}}{2} \, e^{\ell_{p}^{2} \cK} \, \cG^{i \bar{\jmath}} \, \cD_{i} \cW \, \cD_{\bar{\jmath}} \overline{\cW} \, \big( \bbP_{L} \varepsilon \big) \, \\
& = - \frac{\ell_{p}^{3}}{2} \, e^{\ell_{p}^{2}  \frac{\cK}{2}} \, \cD_{i} \cW \, \frac{1}{\ell_{p}} \gamma^{\mu} \partial_{\mu} \phi^{i} \, \big( \bbP_{R} \varepsilon \big) + \frac{\ell_{p}^{2}}{2} \, V_{+} \, \big( \bbP_{L} \varepsilon \big) \, ,
\end{align*}
where in the second term we recognise the positive definite part of the scalar potential, denoted by $V_+$. In the first term the complex quantity 
\begin{align*}
\frac{\ell_{p}^{3}}{2} \, e^{\ell_{p}^{2}  \frac{\cK}{2}} \, \cD_{i} \cW \, 
\end{align*}
defines, for a fixed value of $\phi^{i}$, a direction in the scalar manifold. After a K\"{a}hler transformation, it can be written as
\begin{align*}
\frac{\ell_{p}^{2}}{2} \, e^{\ell_{p}^{2}  \frac{\cG}{2}} \, \cG_{i} \, .
\end{align*}
We  normalise the direction to a unit vector taking 
\begin{align} \label{sGoldstino directions}
g_{i} = \frac{\cG_{i}}{\sqrt{\cG_{j} \cG^{j}}} \, . 
\end{align}

At this point we would like to point out a slight subtlety concerning the terminology of the Goldstino and sGoldstini. For cosmological purposes, in which one usually 
considers time-dependent scalar fields, the definition of the linear combination of spin-$1/2$ fields which gives the Goldstino is slightly
 different from what is discussed above. This is mainly due to the presence of couplings of the schematic form 
$\bar{\psi} (\partial \phi) \chi$ in (\ref{Kinetic Lagrangian}). In that case a more careful analysis applies which can be found for 
instance in \cite{SUGRABOOK}. Therefore, referring to the $g_{i}$ directions as the sGoldstini is a small abuse of notation in the time-dependent case. 
Nevertheless, these directions can be defined on the scalar manifold as long as supersymmetry is broken and we will use this in what follows.

\subsubsection*{A geometric bound}

In this section we follow the steps of \cite{Covi} and consider the projection of the mass matrix on the direction specified by $g_{i}$. 
For any complex quantity $U_{i}$ with $U_{i} \, \bar{U}^{i} = 1$ we could define two dinstinct real orthonormal directions 
$ (U_{i},\, \bar{U}_{\bar{\imath}})/\sqrt{2}$ and $(i \, U_{i}, \, -i \, \bar{U}_{\bar{\imath}})/\sqrt{2}$. The same could be done with 
the sGoldstino direction $g_{i}$. Consider now the projection of the mass matrix along these directions
\begin{align*}
\frac{1}{2} \, \begin{matr}{cc} g_{i} & g_{\bar{\imath}} \end{matr} \, \begin{matr}{cc} m^{2 \, i}{}_{j} & m^{2 \, i}{}_{\bar{\jmath}} \\ m^{2 \, \bar{\imath}}{}_{j} & m^{2 \, \bar{\imath}}{}_{\bar{\jmath}} \end{matr} \, \begin{matr}{c} g^{j} \\ g^{\bar{\jmath}} \end{matr} \quad , \qquad \frac{1}{2} \, \begin{matr}{cc} - g_{i} & g_{\bar{\imath}} \end{matr} \, \begin{matr}{cc} m^{2 \, i}{}_{j} & m^{2 \, i}{}_{\bar{\jmath}} \\ m^{2 \, \bar{\imath}}{}_{j} & m^{2 \, \bar{\imath}}{}_{\bar{\jmath}} \end{matr} \, \begin{matr}{c} - g^{j} \\ g^{\bar{\jmath}} \end{matr} \, ,
\end{align*}
If we take the averaged sum of these two quantities and normalise it w.r.t.~the potential we are left with 
\begin{align} \label{eta sGoldstino}
\eta_{\rm sG} & \equiv \frac{g^{\bar{\imath}} g^{j} \, \cD_{\bar{\imath}} \cD_{j} V}{\ell^{2}_{p} \, V} = \nonumber \\
& =  \frac{2}{3\gamma} + \frac{4}{\sqrt{3}} \frac{1}{\sqrt{1 + \gamma}} \, \Re\left\{g^{i} \, \frac{\cD_{i} V}{V} \right\} + \frac{\gamma}{1+\gamma} \,\frac{G^{\bar{\imath} j}\cD_{\bar{\imath}} V \, \cD_{j} V}{\ell_{p}^{2} \, V^{2}} - \frac{1 + \gamma}{\gamma} \, \tilde{\cR} \, ,
\end{align}
where we have defined
\begin{align} \label{Gamma}
\gamma = \frac{\ell_{p}^{4} \, V}{3 \, e^{\ell_{p}^{2} \cG}} = \frac{\ell_{p}^{2} \, V}{3 \, |m_{\nicefrac{3}{2}}|^{2}} \, ,
\end{align}
with $m_{3/2}$ being the gravitino mass, $\Re$ denotes the real part and $\tilde{\cR}$ is the sectional curvature related to the plane defined by $g_{i}$ on the scalar manifold
\begin{align} \label{Sectional curvature}
\tilde{\cR} \equiv \frac{\cR_{\bar{\imath} j \bar{k} l} \, g^{\bar{\imath}} g^{j} g^{\bar{k}} g^{l}}{\ell_p^2} \, .
\end{align}
Notice that $\eta_{\rm sG}$ is obtained from the averaged sum of two masses. We will come back to this point in the next section.
In \cite{Covi} $\eta_{\rm sG}$ is used to obtain a bound on the second slow-roll parameter $\eta$ depending on the first 
slow-roll parameter $\epsilon$, $\gamma$ and the sectional curvature $\tilde{\cR}$. In order to get the bound we first notice that, 
for any unit vector $U_{I} = (U_{i} , \, \bar{U}_{\bar{\imath}})/\sqrt{2}$ with $U_{i} \, \bar{U}^{i} = 1$, we have
\begin{align}
\eta \leq \frac{U_{I} \, m^{2 \, I}{}_{J} \, U^{J}}{V} \quad , \qquad \left| \bar{U}^{i} \, \frac{\cD_{i} V}{V} \right| \leq \sqrt{\epsilon} \, . \label{epsilon-inequality}
\end{align}
Combining this information and pluging it into (\ref{eta sGoldstino}), we obtain
\begin{align} 
\label{General bound}
\eta \leq \eta_{\rm sG} \leq  \frac{2}{3\gamma} + \frac{4}{\sqrt{3}} \frac{1}{\sqrt{1 + \gamma}} \, \sqrt{\epsilon} + \frac{\gamma}{1+\gamma} \, \epsilon - \frac{1 + \gamma}{\gamma} \, {\tilde{\cR}} \, .
\end{align}
We will be interested in the last inequality of the chain (\ref{General bound}), namely the one which relates $\eta_{\rm sG}$ to $\epsilon$ and $\tilde{\cR}$. 
This bound is very interesting as it relates the slow-roll parameters to the geometry of the scalar manifold. 
In the next section, after a small summary regarding all the quantities appearing in (\ref{General bound}),  we  analyse their 
inflationary implications.

\section{Inflationary implications}

In this section we discuss the implications of the geometric bound we derived above, (\ref{General bound}). 
In order to do that, we first recap the information contained in this bound and its physical meaning:

\begin{itemize}
\item $\gamma$ is the ratio between the scalar potential and the gravitino mass (\ref{Gamma}). It tells one which is the relative importance 
between the two contributions to the scalar potential. If $\gamma < 0$ the scalar potential is dominated by the negative definite 
contribution. When $\gamma \sim 0$ the two terms are of the same order. Finally when $\gamma > 0$ the supersymmetry breaking  F-terms dominate over the AdS scale, leading to a positive scalar potential.

As we will find below, the bound (\ref{General bound}) turns out to have two regimes. The first one is where $\gamma$ lies in between $0$ and $4/3$, corresponding to a gravitino mass that is above the Hubble scale $H$: $|m_{3/2}|^2 \geq 3 H^2 /4$. The second possibility is when the gravitino mass is below the Hubble scale, corresponding to $\gamma > 4/3$ or in other words $|m_{3/2}|^2 < 3 H^2 / 4$. This is the natural scenario if the gravitino mass during inflation does not differ very strongly from the present gravitino mass, which should be of the order of $1$ TeV in order to address the hierarchy problem. We will assume that this is the case in what follows, and show that the geometric bound poses strong constraints in this regime.

\item $\eta_{\rm sG}$ is the averaged sum of two scalar masses normalised to the value of the potential. If we want to embed effectively
single field inflation in F-term supergravity, one can envision two extreme scenarios. In the first one, the inflaton is not one of the sGoldstino directions. 
In this case, if we want the sGoldstino fields to be spectators during inflation, their masses should be of order $H$  or above 
and hence $\eta_{\rm sG} \gtrsim 1$. 
In the other extreme scenario, the inflaton is one of the sGoldstino directions: this is referred to as sGoldstino inflation (for recent analyses, see e.g.~\cite{AlvarezGaume:2010, *AlvarezGaume:2011a, *AlvarezGaume:2011b, Ana}). Even in this case $\eta_{\rm sG}$ should be of order $1/2$ or larger, because the orthogonal 
sGoldstino field needs to be stabilised along the inflationary trajectory. The general case would be in between these two possibilities, and therefore {\it single field inflation} always requires $\eta_{\rm sG} \gtrsim 1/2$. 

\item $\epsilon$ is the generalisation of the first slow-roll parameter to the case of many scalar fields. It is a measure of the sum of the squared velocity of all the fields. Despite the multi-field generalisation, {\it slow-roll inflation} requires $\epsilon \ll 1$.

\item $\tilde{\cR}$ is the sectional curvature related to the plane identified by the sGoldstino directions. In general the Riemann tensor of a 
manifold is completely specified once all the sectional curvatures are given. For our purposes it is sufficient to say that, if $\tilde{\cR} \sim 1$, 
there are some components of the Riemann tensor which are of order $\ell_{p}^{-2}$ and as a consequence we are dealing with a strongly curved scalar 
manifold. In other words, the scalar kinetic terms in (\ref{Kinetic Lagrangian}) cannot be simply given by
\begin{align*}
- \cG_{i \bar{\jmath}} \, \partial_{\mu} \phi^{i} \, \partial^{\mu} \bar{\phi}^{\bar{\jmath}} \simeq - \sum_{i = 1}^{n} \partial_{\mu} \phi^{i} \, \partial^{\mu} \bar{\phi}^{\bar{\imath}} \, ,
\end{align*}
but one needs to take into account the presence of the K\"{a}hler metric. Therefore {\it canonical kinetic terms} require $\tilde{\cR}=0$.

\end{itemize}

Let us now discuss the inflationary implications of the bound derived above. It will turn out that, for inflationary scenarios with $\gamma > 4/3$, one can only  impose consistently two of the conditions $\{$single field, slow-roll, canonical kinetic terms$\}$ together. On the other hand, for inflationary models with $ 0 < \gamma \le 4/3$, it might be possible to realise the three conditions at the same time in some cases. Let us discuss the three possible consistent combinations.

\subsubsection*{Slow-roll single field inflation}

The first possibility consists of imposing the first two conditions: slow-roll and effective single field inflation. In this case the geometric bound 
\eqref{General bound} becomes 
\be\label{limit1}
   \tilde{\cR} \lesssim \frac{4 - 3 \gamma}{6 (1 + \gamma)} \,,
 \ee
and we see that the sectional curvature of the scalar manifold must be strictly negative for $\gamma>4/3$. In other words, slow-roll and 
single field inflation require us to have non-canonical kinetic terms for the inflaton and all the scalar fields present. Moreover, the 
non-canonical kinetic terms should correspond to a metric whose Riemann curvature has a number of components which are negative and of 
order order one in Planck units. Note that this rules out a number of examples discussed in \cite{Covi}.

The fact that non-canonical terms are required at any point in field space implies that the full inflationary trajectory should extend to 
the point where these terms become relevant -- if this were not the case then inflation should proceed independent of these terms, which we 
know is inconsistent with the bound \eqref{General bound}. Therefore the requirement of non-canonical kinetic terms, with corrections to the metric of order one in Planck units, 
implies that we must have large field inflation. As a consequence, effectively single field and slow-roll cannot be realised in small field F-term inflation. 
Note that this statement on the full inflationary trajectory follows from an analysis of the bound \eqref{General bound} for a single 
point in field space.

There is a small caveat to this statement. Indeed $\tilde{\cR}$ is a specific sectional curvature associated to the plane defined by the sGoldstino fields. By carefully constructing the K\"{a}hler- and super-potential it is possible to obtain an inflationary trajectory along which the inflaton is completely orthogonal to the sGoldstino fields (see e.g.~\cite{Kallosh1, *Kallosh2}). The latter are stabilised and, even being $\tilde{\cR}\ne 0$
still one can obtain canonical kinetic terms for the inflaton allowing for small field inflation. The special features of this model provide an escape from our conclusions. On the other hand, as long as there is a non-negligible overlap between the inflaton and the sGoldstino fields along the inflationary path, our analysis applies. 

Observationally, the consequence of having large field inflation is the prediction that tensor modes can be detectable. The argument proceeds 
via the Lyth bound \cite{LythB}, which relates inflationary trajectories of order one in Planck units to a ratio $r$ between tensor to scalar 
perturbations of percent level. The latter corresponds to observable tensor modes, which are therefore a prediction of F-term inflation.

Furthermore, the implications for the curvature perturbations in inflation with non-standard kinetic terms
have been studied largely in the literature, starting with the work of Garriga-Mukhanov \cite{GM}.
Writing the scalar part of the lagrangian as a general function $P(X,\phi)$,
with $X=\frac12 g^{\mu\nu}\partial_{\mu}\phi\partial_\nu\phi$, we see that the kinetic term for the inflaton  gives rise to a linear function of $X$ in
the present case.
Thus, using the results of \cite{GM}, one sees that the resulting perturbations coincide with the canonical case. In particular, the ``speed of sound" of
the perturbations $c_s$, equals the speed of light. In this case, possible departures from the Gaussian spectrum in the equilateral configuration,
parameterised by $f_{NL}^{eq}\propto 1/c_s^2$ are negligible \cite{eqnG, *CHKS}.
Moreover, as has been shown in \cite{CZ}, non-Gaussianities of the local form $f_{NL}^{loc}$, are suppressed by $1-n_s$ for single field inflation.
Thus in this case, one obtains standard single field predictions for the scalar perturbations.

Finally we note from \eqref{limit1}  that for the other regime with $\gamma  \leq 4/3$, which corresponds to a gravitino mass equal to $|m_{3/2}|^2 \geq 3 H^2 /4$, canonical kinetic terms are possible. 

\subsubsection*{Slow-roll with canonical kinetic terms}

The next possibility is to  impose slow-roll inflation and canonically normalised fields. Thus the bound \eqref{General bound} becomes 
\be\label{limit2}
\eta_{\rm sG}\le \frac{2}{3\gamma} \,.
\ee
This implies that for inflationary models with $\gamma>4/3$, we have to consider multifield inflation with canonical terms for all the fields. 
In this case, large non-Gaussianity can be generated dynamically by inflation due to the interplay of all fields and  large isocurvature 
perturbations. 
 Large non-Gaussianity of the local form $f_{NL}^{loc}$ generated during inflation has been shown to be generically hard to achieve 
(for a review with several references see \cite{BC})
and is very much model dependent. Therefore, without knowledge on the form of the potential, we can only conclude 
that potentially large non-Gaussianities due to multifield dynamics could be generated in these type of models. 

On the other hand, for $\gamma \le 4/3$ one can still realise single field inflation.

\subsubsection*{Single field with canonical kinetic terms}
The last possible combination is to impose effective single field inflation with canonical kinetic terms. In this case, the geometric bound 
\eqref{General bound} translates in a bound for $\epsilon$:
\be\label{limit3}
\sqrt \epsilon \ge \frac{\sqrt{1+\gamma}}{\sqrt{3} \gamma} \left[ -2 + \sqrt{2(1+3\gamma/4)} \right] \,.
\ee
In this case we see that for most values of $\gamma > 4/3$, slow roll inflation cannot be realised. In particular, for $\gamma \gg 1$ one finds $\epsilon \gtrsim 1/2 $.

\section{Conclusions}

In this paper we have analysed the inflationary implications of the sGoldstino bound 
\eqref{General bound}.
The derivation of this bound closely follows \cite{Covi} and involves the sGoldstini, the two scalar directions that have a special status due to supersymmetry breaking. Whereas the focus in \cite{Covi} was on slow-roll, we extended the analysis with the possible 
requirements of effectively single field and/or canonical kinetic terms. 

Remarkably, under the assumption of a sub-Hubble gravitino mass, the combination of the slow-roll and single field imposes a very strong constraint on F-term inflation. The curvature of the K\"ahler 
manifold spanned by the chiral scalars necessarily includes negative components which are order one in Planck units. Only K\"ahler manifolds with this 
property satisfy the necessary but not sufficient condition for slow-roll, single field inflation. This rules out many of the examples considered in 
the literature, see e.g.~\cite{Covi}. Moreover, as discussed in the previous section, this automatically implies that the full inflationary trajectory 
will be in the large field class. A consequence is the generation of observable tensor modes in the polarisation of the CMB.

In the very recent and related paper \cite{Ana}, a general analysis has been performed of sGoldstino inflation, where it is assumed that the inflaton 
coincides with one of the sGoldstini directions. It is very interesting to compare the findings of that paper to our results above. First of all, 
we have verified that the two explicit trajectories presented in section 3.3 of \cite{Ana} not only satisfy the bound \eqref{General bound}, but actually 
saturate it. The latter can be understood from the second inequality of \eqref{epsilon-inequality}, which reduces to an equality in the case of one chiral 
multiplet. This is a general feature of sGoldstino inflation. Secondly, in the set-up discussed in \cite{Ana}, it is claimed that large field inflation is 
impossible. Combined with our geometric bound, this would completely rule out single field, slow-roll inflation in such a scenario. It would clearly be 
worthwhile to deepen our understanding of these restrictions arising from the sGoldstino sector.

\section*{Acknowledgements}
We thank  R.~Kallosh,  P.~ Ortiz and G.~Tasinato for interesting and stimulating discussions.
The authors are supported by a VIDI grant from the Netherlands Organisation for Scientific Research (NWO). 

\providecommand{\href}[2]{#2}\begingroup\raggedright\endgroup


\begin{thebibliography}{10}

\bibitem{Guth:1980zm}
A.~H. Guth,  {\em {The Inflationary Universe: A Possible Solution to the
  Horizon and Flatness Problems}}, Phys.Rev. {\bf D23} (1981)
347--356.

\bibitem{Linde:1981mu}
A.~D. Linde,  {\em {A New Inflationary Universe Scenario: A Possible Solution
  of the Horizon, Flatness, Homogeneity, Isotropy and Primordial Monopole
  Problems}}, Phys.Lett. {\bf B108} (1982)
389--393.

\bibitem{WMAP7}
{\bf WMAP Collaboration} Collaboration, E.~Komatsu {\em et al.},  {\em
  {Seven-Year Wilkinson Microwave Anisotropy Probe (WMAP) Observations:
  Cosmological Interpretation}}, Astrophys.J.Suppl. {\bf 192} (2011) 18
  [\href{http://www.arXiv.org/abs/1001.4538}{{\tt 1001.4538}}].

\bibitem{Planck}
{\bf Planck Collaboration} Collaboration, P.~Ade {\em et al.},  {\em {Planck
  Early Results: The Planck mission}},
\href{http://www.arXiv.org/abs/1101.2022}{{\tt 1101.2022}}.

\bibitem{DbraneIn}
L.~McAllister and E.~Silverstein,  {\em {String Cosmology: A Review}},
  Gen.Rel.Grav. {\bf 40} (2008) 565--605
[\href{http://www.arXiv.org/abs/0710.2951}{{\tt 0710.2951}}].

\bibitem{LV}
V.~Balasubramanian, P.~Berglund, J.~P. Conlon and F.~Quevedo,  {\em
  {Systematics of moduli stabilisation in Calabi-Yau flux compactifications}},
  JHEP {\bf 0503} (2005) 007
[\href{http://www.arXiv.org/abs/hep-th/0502058}{{\tt hep-th/0502058}}].

\bibitem{CQ}
J.~P. Conlon and F.~Quevedo,  {\em {Kahler moduli inflation}}, JHEP {\bf 0601}
  (2006) 146
[\href{http://www.arXiv.org/abs/hep-th/0509012}{{\tt hep-th/0509012}}].

\bibitem{DeWolfe}
O.~DeWolfe, A.~Giryavets, S.~Kachru and W.~Taylor,  {\em {Type IIA moduli
  stabilization}}, JHEP {\bf 07} (2005) 066
[\href{http://www.arXiv.org/abs/hep-th/0505160}{{\tt hep-th/0505160}}].

\bibitem{Hertzberg}
M.~P. Hertzberg, S.~Kachru, W.~Taylor and M.~Tegmark,  {\em {Inflationary
  Constraints on Type IIA String Theory}}, JHEP {\bf 12} (2007) 095
[\href{http://www.arXiv.org/abs/0711.2512}{{\tt 0711.2512}}].

\bibitem{mcallister}
N.~Agarwal, R.~Bean, L.~McAllister and G.~Xu,  {\em {Universality in D-brane
  Inflation}}, JCAP {\bf 1109} (2011) 002
  [\href{http://www.arXiv.org/abs/1103.2775}{{\tt 1103.2775}}].

\bibitem{GR1}
M.~Gomez-Reino and C.~A. Scrucca,  {\em {Locally stable non-supersymmetric
  Minkowski vacua in supergravity}}, JHEP {\bf 05} (2006) 015
[\href{http://www.arXiv.org/abs/hep-th/0602246}{{\tt hep-th/0602246}}].

\bibitem{GR2}
M.~Gomez-Reino and C.~A. Scrucca,  {\em {Constraints for the existence of flat
  and stable non- supersymmetric vacua in supergravity}}, JHEP {\bf 09} (2006)
  008
[\href{http://www.arXiv.org/abs/hep-th/0606273}{{\tt hep-th/0606273}}].

\bibitem{Louis}
M.~Gomez-Reino, J.~Louis and C.~A. Scrucca,  {\em {No metastable de Sitter
  vacua in N=2 supergravity with only hypermultiplets}}, JHEP {\bf 02} (2009)
  003
[\href{http://www.arXiv.org/abs/0812.0884}{{\tt 0812.0884}}].

\bibitem{Covi}
L.~Covi, M.~Gomez-Reino, C.~Gross, J.~Louis, G.~A. Palma {\em et al.},  {\em
  {Constraints on modular inflation in supergravity and string theory}}, JHEP
  {\bf 0808} (2008) 055
[\href{http://www.arXiv.org/abs/0805.3290}{{\tt 0805.3290}}].

\bibitem{Ana}
A.~Achucarro, S.~Mooij, P.~Ortiz and M.~Postma,  {\em {Sgoldstino inflation}},
\href{http://www.arXiv.org/abs/1203.1907}{{\tt 1203.1907}}.

\bibitem{SUGRABOOK}
D.~Freedman and A.~Van~Proeyen,  {\em {Supergravity}}, Cambridge University
  Press (2012) to be published.

\bibitem{Cremmer:1982wb}
E.~Cremmer, S.~Ferrara, L.~Girardello and A.~Van~Proeyen,  {\em {Coupling
  Supersymmetric Yang-Mills Theories to Supergravity}}, Phys.Lett. {\bf B116}
  (1982)
231.

\bibitem{AlvarezGaume:2010}
L.~Alvarez-Gaume, C.~Gomez and R.~Jimenez,  {\em {Minimal Inflation}},
  Phys.Lett. {\bf B690} (2010) 68--72
[\href{http://www.arXiv.org/abs/1001.0010}{{\tt 1001.0010}}].

\bibitem{AlvarezGaume:2011a}
L.~Alvarez-Gaume, C.~Gomez and R.~Jimenez,  {\em {A Minimal Inflation
  Scenario}}, JCAP {\bf 1103} (2011) 027
[\href{http://www.arXiv.org/abs/1101.4948}{{\tt 1101.4948}}].

\bibitem{AlvarezGaume:2011b}
L.~Alvarez-Gaume, C.~Gomez and R.~Jimenez,  {\em {Phenomenology of the minimal
  inflation scenario: inflationary trajectories and particle production}},
\href{http://www.arXiv.org/abs/1110.3984}{{\tt 1110.3984}}.

\bibitem{LythB}
D.~H. Lyth,  {\em {What would we learn by detecting a gravitational wave signal
  in the cosmic microwave background anisotropy?}}, Phys.Rev.Lett. {\bf 78}
  (1997) 1861--1863
[\href{http://www.arXiv.org/abs/hep-ph/9606387}{{\tt hep-ph/9606387}}].

\bibitem{GM}
J.~Garriga and V.~F. Mukhanov,  {\em {Perturbations in k-inflation}}, Phys.
  Lett. {\bf B458} (1999) 219--225
[\href{http://www.arXiv.org/abs/hep-th/9904176}{{\tt hep-th/9904176}}].

\bibitem{eqnG}
M.~Alishahiha, E.~Silverstein and D.~Tong,  {\em {DBI in the sky}}, Phys. Rev.
  {\bf D70} (2004) 123505
[\href{http://www.arXiv.org/abs/hep-th/0404084}{{\tt hep-th/0404084}}].

\bibitem{CHKS}
X.~Chen, M.-x. Huang, S.~Kachru and G.~Shiu,  {\em {Observational signatures
  and non-Gaussianities of general single field inflation}}, JCAP {\bf 0701}
  (2007) 002
[\href{http://www.arXiv.org/abs/hep-th/0605045}{{\tt hep-th/0605045}}].

\bibitem{CZ}
P.~Creminelli and M.~Zaldarriaga,  {\em {Single field consistency relation for
  the 3-point function}}, JCAP {\bf 0410} (2004) 006
[\href{http://www.arXiv.org/abs/astro-ph/0407059}{{\tt astro-ph/0407059}}].

\bibitem{BC}
C.~T. Byrnes and K.-Y. Choi,  {\em {Review of local non-Gaussianity from
  multi-field inflation}}, Adv. Astron. {\bf 2010} (2010) 724525
[\href{http://www.arXiv.org/abs/1002.3110}{{\tt 1002.3110}}].

\bibitem{Kallosh1}
R.~Kallosh, A.~Linde and T.~Rube,  {\em {General inflaton potentials in supergravity}}, Phys.Rev. {\bf D83} (2011) 043507
[\href{http://www.arXiv.org/abs/1011.5945}{{\tt 1011.5945}}].

\bibitem{Kallosh2}
R.~Kallosh, A.~Linde, K.~A.~Olive and T.~Rube,  {\em {Chaotic inflation and supersymmetry breaking}}, Phys.Rev. {\bf D84} (2011) 083519
[\href{http://www.arXiv.org/abs/1106.6025}{{\tt 1106.6025}}].

\end{thebibliography}
\end{document}